# Electric field and current induced electroforming modes in NbO$_x$

Sanjoy Kumar Nandi[1]*, Shimul Kanti Nath[1], Assaad E. El-Helou[2], Shuai Li[1,3], Thomas Ratcliff[1], Mutsunori Uenuma[4], Peter E Raad[2], Robert G Elliman[1]*

[1]Department of Electronic Materials Engineering, Research School of Physics, The Australian National University, Canberra ACT 2601, Australia
[2]Department of Mechanical Engineering, Southern Methodist University, Texas, Dallas, USA
[3]Unité Mixte de Physique, CNRS, Thales, Université Paris-Sud, Université Paris-Saclay, Palaiseau, France
[4]Nara Institute of Science and Technology (NAIST), Information Device Science Laboratory
8916-5, Takayamacho, Ikoma, Nara 630-0192 Japan

E-mail: sanjoy.nandi@anu.edu.au (S. K. Nandi)
E-mail: rob.elliman@anu.edu.au (R. G. Elliman)

Keywords: Electroforming, current bifurcation, non-volatile memory, negative differential resistance, niobium oxide

## ABSTRACT

Electroforming is used to initiate the memristive response in metal/oxide/metal devices by creating a filamentary conduction path in the oxide film. Here we use a simple photoresist-based detection technique to map the spatial distribution of conductive filaments formed in Nb/NbO$_x$/Pt devices, and correlate these with current-voltage characteristics and in-situ thermoreflectance measurements to identify distinct modes of electroforming in low and high conductivity NbO$_x$ films. In low conductivity films the filaments are randomly distributed within the oxide film, consistent with a field-induced weakest-link mechanism, while in high conductivity films they are concentrated in the center of the film. In the latter case the current-voltage characteristics and in-situ thermoreflectance imaging show that electroforming is associated with current bifurcation into regions of low and high current density. This is supported by finite element modelling of the current distribution and shown to be consistent with predictions of a simple core-shell model of the current distribution. These results clearly demonstrate two distinct modes of electroforming in the same materials system and show that the dominant mode depends on the conductivity of the film, with field-induced electroforming dominant in low conductivity films and current-bifurcation induced electroforming dominant in high conductivity films. Finally, we demonstrate S-type and snap-back negative differential resistance in the high conductivity films and explain this behavior in terms of two-zone model.

1. Introduction

Memristive-switching in metal/oxide/metal devices is of current interest as a basis for non-volatile memory and neuromorphic computing devices [1-4]. However, as-fabricated devices are generally in a high resistance state and require a one-off electroforming step to establish the memristive response [2, 5-6]. This is typically achieved by subjecting the film to a voltage or current stress sufficient to form a filamentary conduction path through the film (i.e. soft dielectric breakdown), a process mediated by the generation, drift and diffusion of atoms and ions in response to the applied electric field and local Joule heating [6-7]. The size, resistance and stability of the resulting filaments depend critically on the forming conditions, and particularly on the maximum forming current and associated temperature rise caused by Joule heating [8-9]. Local Joule heating can also cause compositional or structural changes at the oxide/electrode interface that affect the final state of the electroformed device and its switching characteristics [10-11]. As a consequence, understanding details of the electroforming process is an essential requirement for developing devices with specific memristive characteristics.

The voltage required for filament formation generally scales with film thickness, consistent with a field-induced forming mechanism [12-14]. This is commonly attributed to a high-field electrolytic process in which defects are generated heterogeneously at the oxide/electrode interface by oxidation/reduction reactions and drift and diffuse to create a conduction path between the electrodes [8, 15-16]. However, Kumar and Williams have recently proposed an alternative mode of electroforming in $TaO_x$ based on local thermal decomposition of the oxide film [17]. This was shown to be associated with strong current localisation due to the bifurcation of the current distribution into regions of low and high current density under voltage controlled-testing, and was attributed to thermophoresis of oxygen in the temperature field created by local Joule heating. Similar results have also been observed by the Skowronski group but with some notable differences [18-20]. For example, the latter studies attributed the thermal decomposition of $TaO_x$ films to migration of Ta rather than O, and reported current bifurcation rather than field-bifurcation under current controlled testing [21]. Several other studies have addressed the compositional analysis of the conducting filament using techniques such as conducting atomic formce microscopy (C-AFM)[22], transmission synchrotron x-ray spectromicroscop[17], high-resolution transmission electron microscopy (HRTEM) and electron energy loss spectroscopy (EELS)[23-24]. Given the critical role of electroforming in determining the memristive switching characteristics of metal/oxide/metal devices, understanding the differences between electric-field and thermally-induced electroforming and the role of curent bifurcation in elecroforming is of both scientific and technological interest.

In this study we report for the first time two distinct modes of electroforming in $NbO_x$ and show that the dominant mode depends on the electrical conductivity of the oxide film. This is achieved by correlating results from a recently developed filament detection technique, thermoreflectance imaging, quasi-static current-voltage measurements and finite element modelling. Significantly, we demonstrate current bifurcation under current-controlled testing conditions, and provide a physical basis for understanding how this arises based on a recently developed core-shell model of current transport. Using this model we resolve a recent controversy regarding current bifurcation induced filament formation[17-18]. These results advance the understanding of the electroforming mechanism and provides a basis for predicting the dominant mode of electroforming.

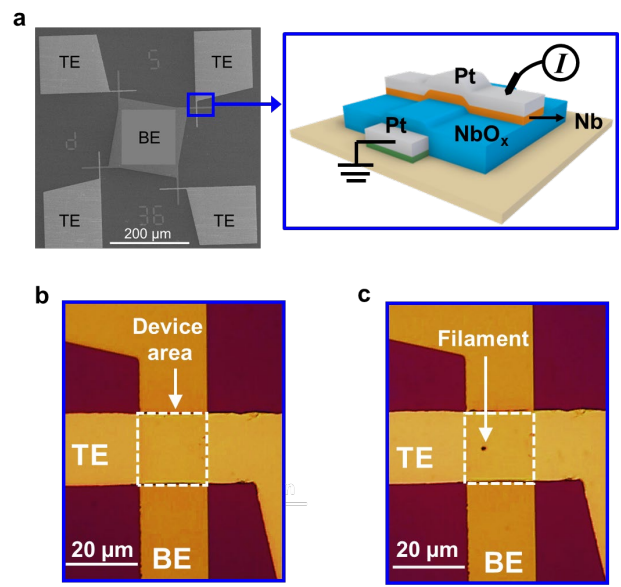

**FIGURE 1: Device structure and filament detection: a)** Scanning electron microscopy image and schematic of the Pt/Nb/NbO$_{2.60}$/Pt device structure. Detail of the device structure is given in the supplementary information Figure S1. **b)** Optical images of a photoresist-coated device before electroforming, and **c)** after electroforming. The dark spot evident in the latter indicates the location of a conducting filament. Figure 1b-c are redrawn from Ref[25].

## 2. Results and Discussion

### 2.1 Filament distribution and modes of electroforming

Our studies were based on Pt/Nb/NbO$_x$/Pt cross-point test structures with electrode dimensions in the range from 2 μm to 20 μm and exploited a recently developed filament detection technique to map the spatial distribution of transient or permanent conductive filaments based on the discoloration of a photoresist layer, as illustrated in Figure 1. Additional insight was provided by in-situ thermoreflectance imaging of device temperature distributions and finite-element modeling of device electrical characteristics and current and temperature distributions.

To highlight a significant outcome of the present study Figures 2a and 2b compare current-controlled electroforming characteristics of 5 μm devices fabricated with low (x=2.60) and high (x=1.92) conductivity NbO$_x$ films. (NB: The relationship between stoichiometry and conductivity is given in the Supplementary Information, Table S1). Both devices exhibit abrupt voltage changes during the forward current sweep that are correlated with local discoloration of the photoresist layer, clearly demonstrating filamentary conduction. However, the reverse current sweeps reveal distinctly different behavior, with the low-conductivity film exhibiting S-type negative differential resistance (NDR) and a permanent change in low-field resistance from several MΩ to ~10 kΩ, and the high-conductivity film exhibiting an abrupt 'snap-back' characteristic with no significant change in resistance. Subsequent current sweeps reproduce the S-type NDR characteristics of the low-conductivity films and the snap-back response of the high-conductivity films in both the forward and reverse sweeps. However, repeated cycling (or high current forming) of the latter eventually produces a permanent resistance change that is evident as a reduction in threshold current during subsequent sweeps (see Supplementary information Figure S4). i.e. The device continues to exhibit a hysteretic snap-back characteristic once a permanent filament is formed but at a lower onset current. The threshold voltage distributions for the low and high conductivity films have also shown distinct differences, with the distribution for the former being narrower and more symmetrical than that of the latter (shown in Supplementary Information Figure S5).

The corresponding filament distributions for these films are shown in Figures 2c and 2d, where each point represents the location of a single filament measured in a separate device. Comparing these figures reveals a dramatic difference in the filament distributions. For the low conductivity films a significant fraction (~80 %) of filaments form at the electrode edges while those within the oxide film are randomly distributed. These observations are consistent with a field-induced, weakest-link model of filament formation based on the drift and diffusion of oxygen vacancies, as illustrated schematically in Figure 2e [5, 9]. Within this framework, the concentration of filaments at the electrode edges reflects variations in the film thickness and local electric field in this region [25]. In contrast, the current flows uniformly (no weak spots) in the high-conducting films and the filament distribution is localized around the center of the device and no filaments are detected at the electrode edges even though the electrode geometry is similar to that for the low-conductivity films. (We note, however, that the filament distribution in these high-conductivity films is very sensitive to the uniformity of the oxide film and the roughness of the bottom electrode.) As we show below, this is consistent with the formation of a transient conductive filament, such as that illustrated in Figure 2f.

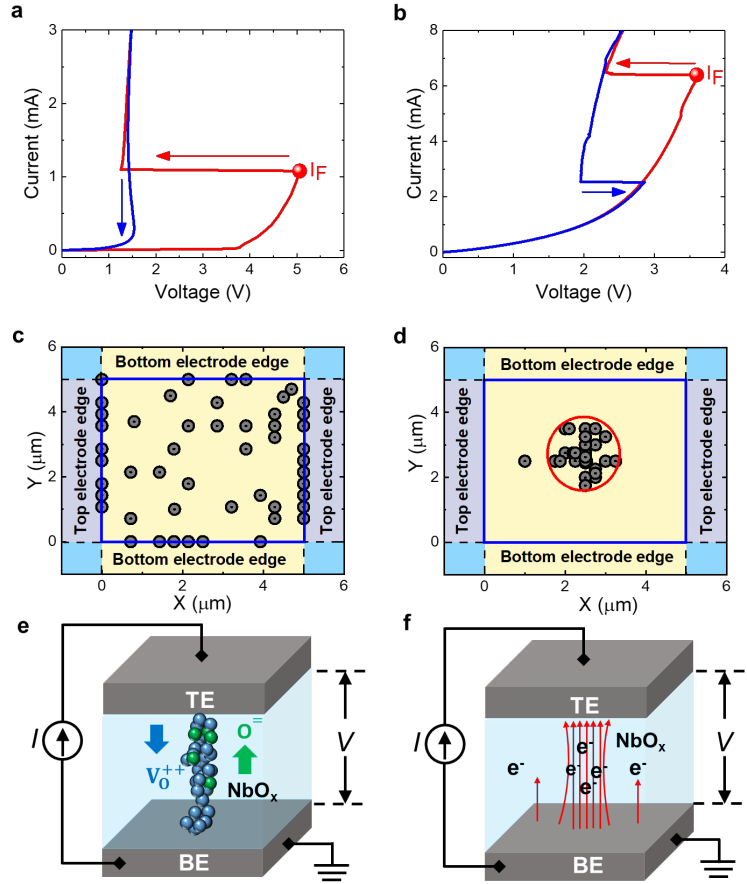

**FIGURE 2**: **Electroforming and filament distributions: a)** Electroforming characteristics of a 5 μm Pt/Nb/NbO$_{2.6}$/Pt device, **b)** electroforming characteristics of a 5 μm Pt/Nb/NbO$_{1.92}$/Pt device, **c)** shematic representation of filament distribution in 5 μm × 5 μm cross-point device with low-conductivity NbO$_x$ (x=2.60) films, **d)** schematic representation of filament distribution in 5 μm × 5 μm cross-point device with high-conductivity NbO$_x$ (x=1.92) films, **e)** schematic of oxygen vacancy filament formed by the field-induced generation, drift and diffusion of oxygen vacancies, and **f)** schematic of a transient current filament due to current bifurcation. Note that each dot in (**c-d**) represents single device and the filament distributions represent 75 separate measurements for each film.

## 2.2 In-situ temperature distribution during current bifurcation

The in-situ thermoreflectance measurements in Figure 3 compare I-V characteristics and temperature distributions of a 10 μm x 10 μm device with a high-conductivity NbOx (x=1.99) film during electroforming. The in-situ I-V characteristics show a snap-back response at a threshold current of ~5.9 mA, consistent with the results reported in Figure 2b. For currents below this forming threshold the temperature increases monotonically with current and has a broad distribution that reflects the boundary conditions imposed by the device geometry. However, at the onset of forming the distribution narrows abruptly and the maximum temperature increases by ~200 K. This produces a local hot spot with a FWHM of ~1.5 μm and is accompanied by a ~50 K temperature reduction in the region surrounding the filament that is consistent with the redistribution of current from regions of low to high current density. These results clearly demonstrate the correlation between the abrupt voltage change and current bifurcation in high conductivity films. It is important to note that the temperature distribution at the top electrode surface is expected to be much broader than the actual width of the conductive filament due to the high thermal conductivity of the metal eletrode. Indeed, back-scattered electron imaging of the oxide surface show a filamentary core of approximately 300 nm diameter (see inset in Figure 3c). Interestingly, these image also reveal that the core is surounding by region of lower average atomic number, consistent with the formation of an oxygen rich halo around a central conductive core, as reported by others [21, 26].

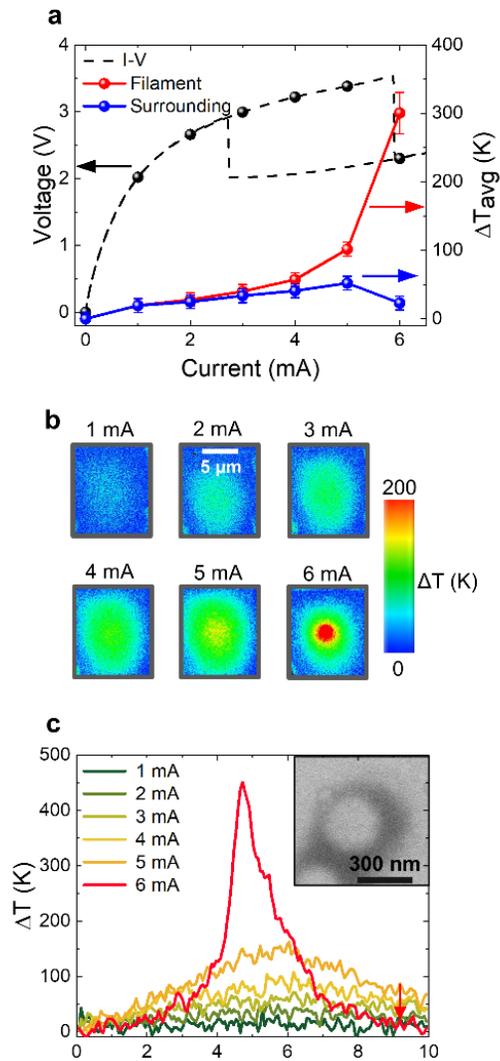

**FIGURE 3: In-situ thermoreflectance measurements: a)** In-situ (points) and ex-situ (line) current-voltage characteristics of a 10 μm × 10 μm cross-point device with a high-conductivity $NbO_x$ (x=1.92) film, and the temperature of the filamentary and surrounding regions of the film (shown by circles in panel b), **b)** 2D temperature maps of the top electrode surface for different device currents, **c)** Temperature profile through the filamentary region for different device currents. Inset is a back-scattered electron image of the filamentary region in the oxide film (i.e. after removing the top electrode).

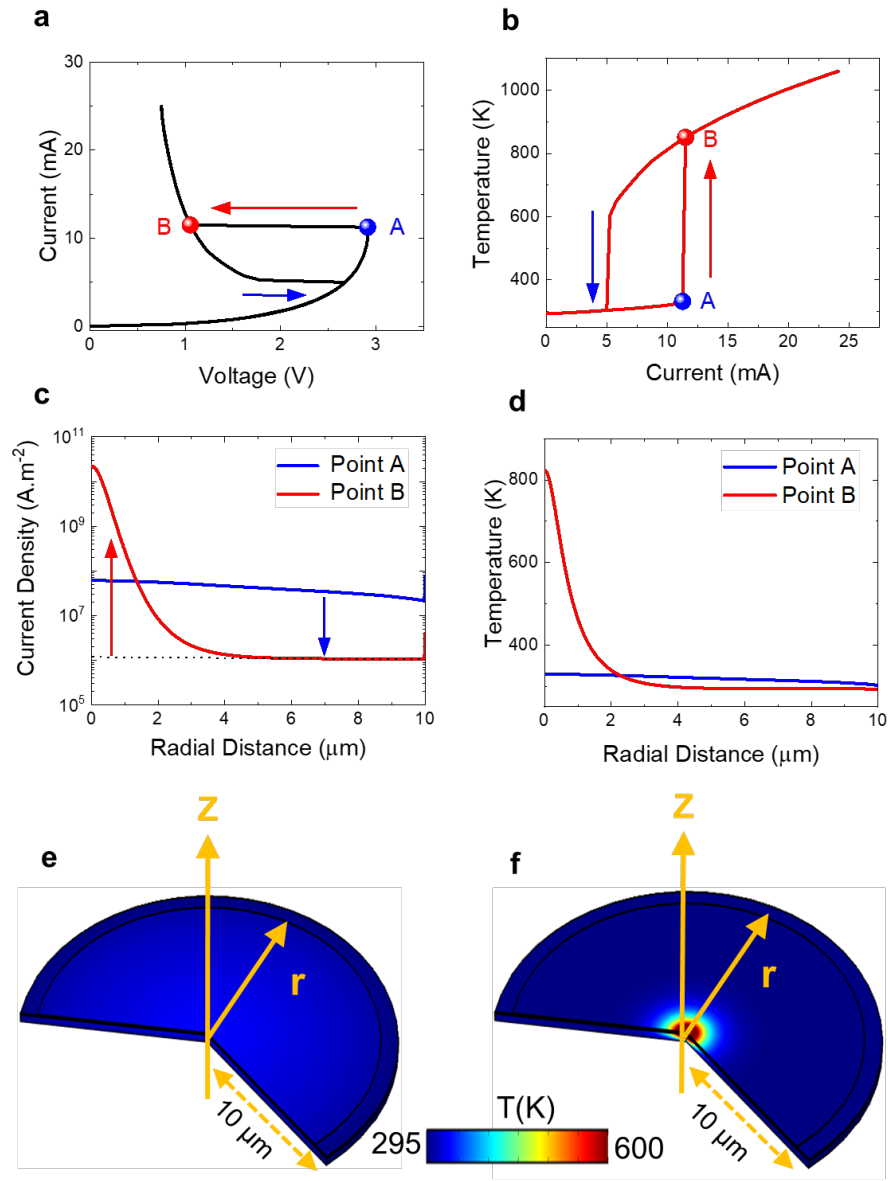

**FIGURE 4: Finite element simulation: a)** Current-voltage characteristics during bidirectional current sweep, **b)** Maximum temperature in oxide film as a function of device current, **c)** current density distribution before and after bifurcation, **d)** temperature distribution before and after bifurcation. 3D Temperature distribution **e)** before (point A) and **f)** after current bifurcation (point B) temperature distributions after current bifurcation.

*2.3 Electro-thermal model of current bifurcation*

A finite-element model of the bifurcation process was used to provide a comparison with the experimental data in Figures 2 and 3 and support the above interpretation. The device structure was represented by a 2D axisymmetric model in which the electrical conductivity of the oxide film was governed by Poole-Frenkel conduction. The current-voltage characteristics and associated current and temperature distributions were then calculated by self-consistently solving the heat transfer and current continuity equations, including the effect of Joule heating. (Further details of the model and associated parameters are given in the Supplementary Information) Selected results from the simulation are summarized in Figure 4 and highlight several important points. Specifically, the simulation reproduces the snap-back characteristics observed in Figure 2a without imposing materials specific properties or inhomogeneity, and predicts the formation of high current density 'filaments' at the center of the device structure, consistent with the filament distribution reported in Figure 2d. The latter is the result of current redistribution from regions of low to high current density, and produces a corresponding change in the temperature of the filament and surrounding film as observed experimentally in Figure 2. Finally, the model predicts area-dependent NDR characteristics, with small devices exhibiting S-type NDR characteristics and larger ones exhibiting snap-back characteristics, consistent with experimental observations and the simulation results of Goodwill et al. [18].

*2.4 Bifurcation and NDR response*

The above results explain the origin of the NDR responses observed during forward current sweeps but do not explain why the low conductivity films exhibit S-type NDR and the high conductivity films exhibit snap-back NDR during reverse current sweeps (see Figure 2). It is important to note here that reverse sweep is equivalent to the subsequent sweep immediately after the electroforming step. For this we draw on a simple two-zone, lumped element model of the current distribution that has been shown to explain such behavior [27-28]. This assumes that highly localized current distributions, such as those depicted in Figures 3 and 4, can be represented by a high-temperature memristive core and a low temperature resistive shell, as shown in Figure 5a. The shell resistor then acts as a current divider that enables current redistribution in response to the reduction in core resistance created by Joule heating. A snap-back response is then observed if the core temperature exceeds the threshold required for negative differential resistance and the shell resistance, $R_S$, is less than the maximum negative differential resistance of the filamentary core, $R_{NDR}$ (i.e. $R_S < R_{NDR}$) as shown in Figure 5a [27]. If the latter condition is not met the I-V characteristics are dominated by the NDR response of the core and the device exhibits S-type NDR. The reverse current sweeps of Figure 2 can then be understood by considering the effect of the film conductivity on the shell resistance. Specifically, the S-type NDR response of the low conductivity film can be understood by assuming that electroforming creates a permanent filament with NDR characteristics and that the surrounding low-conductivity film has a resistance such that $R_S > R_{NDR}$. The I-V characteristic of the device during the reverse current sweep is then dominated by the NDR response of the filamentary core region. In contrast, the snap-back characteristics of the high conductivity film can be understood by assuming that a transient filament forms as a result of current bifurcation and that the surrounding high-conductivity film has a resistance such that $R_S < R_{NDR}$. In this case, the snap-back response is expected whether or not a permanent filament is formed by the bifurcation process.

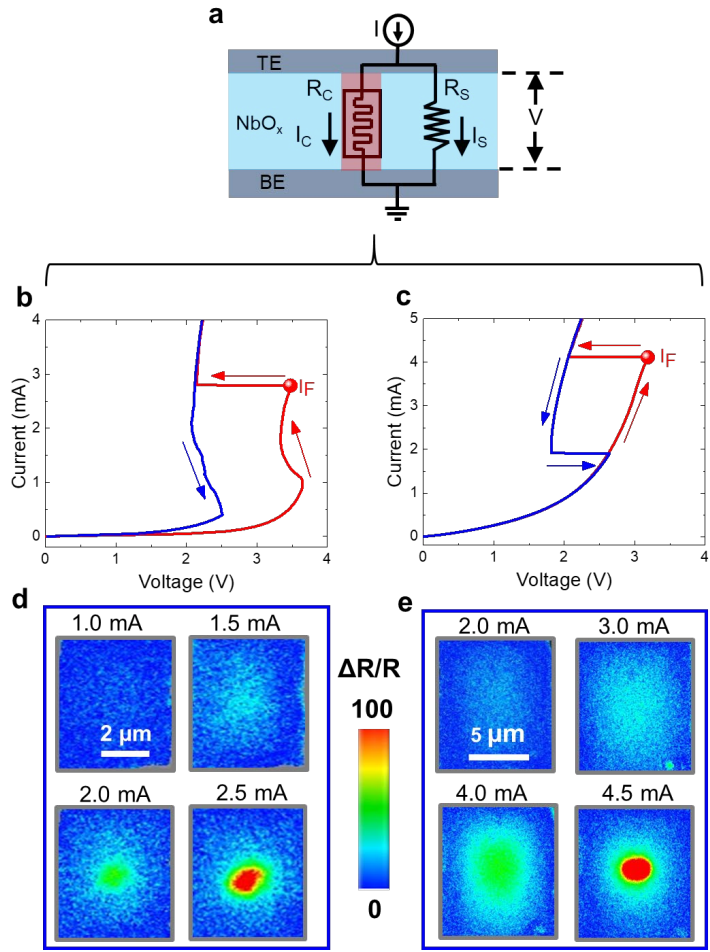

**FIGURE 5**: **Core-shell model and current bifurcation: a)** Schematic of the core-shell model of negative differential resistance device, **b)** electroforming characteristics of a 5 μm × 5 μm device with a NbO$_x$ (x=2.22) film, **c)** electroforming characteristics of a 10 μm × 10 μm device with a NbO$_x$ (x=2.22), **d)** In-situ thermoreflectance (ΔR/R) maps of a 5 μm × 5 μm device during electroforming, and **e)** In-situ thermoreflectance (ΔR/R) maps of a 10 μm × 10 μm device during electroforming. Note: Thermoreflectance mapping was performed on SiN$_x$/Pt/Nb/NbO$_{2.22}$/Pt device structures.

As a final example of how this model can be used to understand diverse current-voltage characteristics we consider the characteristics in Figures 5b and 5c. These show bi-directional forming characteristics of 5 μm and 10 μm cross-point devices with NbO$_x$ stoichiometry x=2.22. In this case the film conductivity is the same for both devices so differences in shell resistance are, in principle, determined by the device area rather than the film conductivity. The smaller area device exhibits S-type NDR during the reverse current sweep, consistent with the requirements that $R_S > R_{NDR}$, while the larger area device exhibits a snap-back response, consistent with the requirement that $R_S < R_{NDR}$. However, while the I-V characteristics of the smaller device is similar to that of the low conductivity film in Figure 2a, the spatial distribution of filaments is localized around the center of the film, similar to that observed for high conductivity films in Figure 2b, a point that is confirmed by the thermoreflectance images in Figures 5d and 5e. So while the spatial distribution of filaments is consistent with current bifurcation and the condition $R_S < R_{NDR}$, the reverse current sweep exhibits S-type NDR, consistent with $R_S > R_{NDR}$. To resolve this conundrum within the limitations of the core-shell model we need to relax the assumption of a fixed shell resistance and consider the more general case where it is temperature dependent. Indeed, the most general two-zone model treats the core and shell regions as parallel memristors and has been shown to reproduce a diverse range of compound NDR characteristics [27]. Here we simply need to note that the temperature of the shell region increases with increasing current prior to filament formation and decreases once the filament forms due to a reduction in local current density (See Figure 3 and 4) and that this produces a concomitant change in shell resistance. Taking this into account the data of Figure 5b can be interpreted as follows: During the forward current sweep the device temperature increases with increasing current such that the peak of the temperature distribution exceeds the threshold for NDR while the resistance of the surrounding shell remains too high to initiate current bifurcation (i.e. $R_S > R_{NDR}$). As a consequence the device exhibits S-type NDR. However, as the current continues to increase, the shell resistance decreases to the point that $R_S < R_{NDR}$ and the current bifurcates into regions of high and low current density. This produces a permanent filament at the point of peak current density and temperature, consistent with the filament distribution. Once the filament is formed, the temperature of the shell decreases and the shell resistance increases such that $R_S > R_{NDR}$ and the device exhibits S-type NDR during the reverse current sweep, as expected. The magnitude of the expected resistance change can be estimated from the equation for Poole-Frenkel conduction (assuming a trap energy level in the range 0.25 – 0.5 eV) and the change in shell temperature (~50 K) and corresponds to a factor of 4 – 15, clearly sufficient to satisfy the above assumptions. This example highlights the power of the core-shell model and filament mapping for understanding complex electroforming and NDR characteristics in metal-oxide-metal device structures.

3. Summary and conclusions

The above results have identified two distinct electroforming mechanisms and shown that post-forming CC-NDR characteristics can be explained within the framework of a previously reported two-zone, core-shell model [27]. While the model involves a major simplification of the problem it does provide unique insight into the underlying physical mechanisms and a simple understanding of experimentally observed characteristics. Significantly, it also has the potential to resolve apparently contradictory results. For example, whereas we have clearly demonstrated current bifurcation and filament formation under current-controlled testing, Kumar and Williams[17] only observed such behavior under voltage controlled operation. While it is tempting to draw on the theoretical analysis of Ridley [29] to help resolve this issue it is important to note that his analysis only considered bifurcation due to electronics instabilities and ignored effects due to temperature variations. As a consequence, his analysis does not apply to the studies in question. However, the observed differences can be understood within the framework of the core-shell model by accounting for the role of the shell resistance, $R_S$, and assuming that the devices used by Kumar and Williams [17] satisfied the condition $R_S > R_{NDR}$ while those used in our study satisfied the condition $R_S < R_{NDR}$. Under these conditions, the current distribution in the former would be expected to constrict gradually while that in the latter would be expected to bifurcate into regions of high and low current density, as observed. While further work is required to confirm this proposition, it clearly demonstrates the potential of the core-shell model for understanding disparate behaviour. It is also interesting to note that within the framework this model, the shell resistance effectively acts as a parallel load resistor for a current source so that the distinction between current-control and voltage control becomes somewhat arbitrary.

In conclusion, we have used quasi-static I-V characteristics, filament mapping, thermoreflectance imaging and finite element modelling to investigate current bifurcation and electroforming and their effect on the CC-NDR characteristics of $NbO_x$-based devices. Two distinct electroforming modes were identified and classified as field-induced or thermally-induced based on their characteristics and dependencies. Field induced electroforming was observed in low-conductivity films and was characterized by a random distribution of filaments within the oxide film and heterogeneity due to material and device inhomogeneity, whereas thermally-induced electroforming was observed in high-conductivity films and was characterized by a highly localized filament distribution at the center of the film, the region of peak temperature. The former was found to be typical of field-induced dielectric breakdown in stoichiometric oxide films while the latter was found to represent a new mode of electroforming. This was mediated by a current bifurcation process in which the current distribution separated into regions of low and high current density and produced a highly localized temperature distribution in the film. Bifurcation produced a sudden drop in voltage under current-controlled testing due to the increase in film conductivity caused by local Joule heating but only created a permanent material change under repeated testing or high-current operation. The post-forming CC-NDR characteristics of devices were also found to depend on device and film parameters, with devices formed by current bifurcation typically exhibiting abrupt snap-back NDR characteristics and those undergoing field-induced forming typically exhibiting S-type NDR characteristics. An example of more complex electroforming and NDR characteristics based on a combination of S-type and snap-back characteristics was also presented. Significantly, all of the reported forming and NDR characteristics were shown to be consistent with the understanding provided by a simple two-zone lumped element model of the device current distribution.

4. Experimental Details

Au/Nb/NbO$_x$/Pt cross-point test structures with electrode dimensions ranging from 2 μm × 2 μm to 10 μm × 10 μm were fabricated using standard photolithographic processing, as shown in Figure 1(a) [25]. The bottom electrode consisted of a 5 nm thick Nb wetting-layer and a 25 nm-thick Pt layer deposited sequentially by e-beam evaporation on a thermally oxidized Si (100) wafer with a 300 nm thick oxide layer. A 45 nm NbO$_x$ (2.60±0.5) dielectric layer was then deposited over the entire wafer, including the lithographically defined bottom electrode, using RF sputtering from an Nb$_2$O$_5$ target. The sub-stoichiometric NbO$_x$ (x=1.92±0.04 to 2.22±0.03) films were deposited using DC sputtering at different pressure in Ar/O$_2$ ambient. The details of the deposition conditions are given in Supplementary Information Table S1. Top electrodes, consisting of a 5 nm-thick Nb layer and a 25 nm Pt protective layer, were then deposited by e-beam evaporation. The oxide layer covering the bottom contact pads was removed by etching to provide direct electrical contact to the pad.

As-deposited NbO$_x$ films were analyzed by grazing incident-angle X-ray diffraction (GIAXRD) and Rutherford backscattering spectroscopy (RBS) to determine their structure and composition, respectively. Electrical measurements were performed using an Agilent B1500A semiconductor parameter analyzer attached to a Signatone probe station (S-1160). All the measurements were executed at room temperature in atmospheric condition by applying current on the top electrode, while the bottom electrode was grounded.

The location of filamentary conduction paths was determined from the discoloration of a photoresist layer during electroforming process [25]. As an example, Figs. 1(b) and 1(c) compare images of a cross-point device before and after electroforming. The dark spot evident in the image of the post-formed device confirms filament formation and clearly identifies the filament location.

Thermal imaging of devices during electroforming was performed using TMX T°Imager®, a camera-based thermoreflectance imaging system [30]. The thermoreflectance method is based on the temperature dependent optical properties of reflective surface materials, and is non-contact and non-destructive (see Supplementary Information Figure S7). The method uses visible light illumination, provides deep submicron resolutions (0.1 - 0.3 µm) well beyond what is possible with other imaging techniques, such as infrared (3-10 µm), and is therefore well suited for the wide range of materials present in microelectronic devices. The temperature-reflectance relation, represented by the relation $\frac{\Delta R}{R} = \left(\frac{1}{R}\frac{\partial R}{\partial T}\right)\Delta T = C_{TR}\,\Delta T$, where $\Delta R$ is the change in reflectance, $\Delta T$ is the change in temperature, and $C_{TR}$ is the thermoreflectance co-efficient for the top electrode material. Additional in-situ temperature mapping for electroforming process for stoichiometric NbO$_x$ is performed in infrared microscopy (see Supplementary Information Figure S8).

Supporting Information

- Table listing deposition conditions of NbO$_x$ films

- Optical images of different processing steps of device fabrication and additional images of filament locations

- Example of a Rutherford back-scattered spectrum obtained from DC reactively sputtered $NbO_x$ thin film

- Electroforming and subsequent NDR characteristics; and forming voltage distributions

- Finite element modelling of the current bifurcation induced electroforming process

- Experimental setup of thermoreflectance and infrared microscopy

Acknowledgements

This work was partly funded by the Australian Research Council (ARC) and Varian Semiconductor Equipment/ Applied Materials through an ARC Linkage Project Grant: LP150100693. We would like to acknowledge access to NCRIS facilities at the ACT node of the Australian National Fabrication Facility (ANFF) and the Australian Facility for Advanced ion-implantation Research (AFAiiR).

# Supporting Information

**Electric field and current induced electroforming modes in NbO$_x$**

Sanjoy Kumar Nandi[1]*, Shimul Kanti Nath[1], Assaad E. El-Helou[2], Shuai Li[1,3], Thomas Ratcliff[1], Mutsunori Uenuma[4], Peter E Raad[2], Robert G Elliman[1]*

[1]Department of Electronic Materials Engineering, Research School of Physics, The Australian National University, Canberra ACT 2601, Australia

[2]Department of Mechanical Engineering, Southern Methodist University, Texas, Dallas, USA

[3]Unité Mixte de Physique, CNRS, Thales, Université Paris-Sud, Université Paris-Saclay, Palaiseau, France

[4]Nara Institute of Science and Technology (NAIST), Information Device Science Laboratory 8916-5, Takayamacho, Ikoma, Nara 630-0192 Japan

E-mail: sanjoy.nandi@anu.edu.au (S. K. Nandi)

E-mail: rob.elliman@anu.edu.au (R. G. Elliman)

## Deposition conditions of NbO$_x$ fims

**Table S1**: Film deposition details including the stoichiometry and thickness of each layer.

| Deposition conditions for NbO$_x$ layers | | | | | | Resisitivity ($\Omega$.m) |
|---|---|---|---|---|---|---|
| Sputter Target | Power (W) | Pressure (mTorr) | Gas flow (Ar/O$_2$) sccm/sccm | Thickness (nm) | x in NbO$_x$ | |
| Nb$_2$O$_5$ | 180 (rf) | 4 | 20/0 | 45 | 2.6±0.05 | 3.5x10$^4$±1.5x10$^3$ |
| Nb | 100 (dc) | 1.5 | 19/1 | 44.3 | 1.92±0.04 | 8±1.5 |
| Nb | 100 (dc) | 2 | 19/1 | 55 | 1.99±0.03 | 15±1 |
| Nb | 100 (dc) | 1.5 | 18.5/1.5 | 57.3 | 2.22±0.04 | 120±7 |

## Steps of the device fabrication

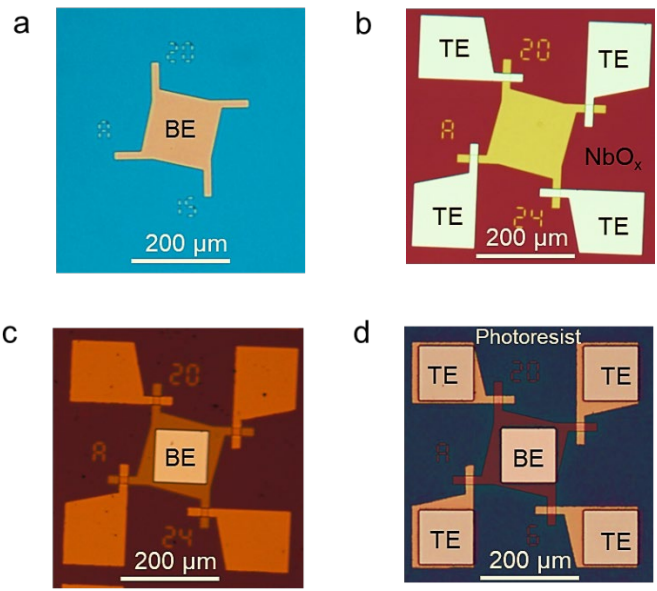

**Figure S1:** Optical micrograph of different processing steps of device fabrication: (a) image of a bottom electrode (BE) after the lift-off process for the deposition of a Pt bottom electrode with a Ti adhesion layer on thermally oxidused Si wafer. (b) Image of four cross-point devices formed with a common bottom electrode (BE) after NbO$_x$ and bottom electrode deposition. (c) The same devices after etching the NbO$_x$ layer to open the BE. (d) Image after the final step, here wafer is coated with photoresist everywhere except the contact pads (BE and TE). Further details can be found in Ref [25]

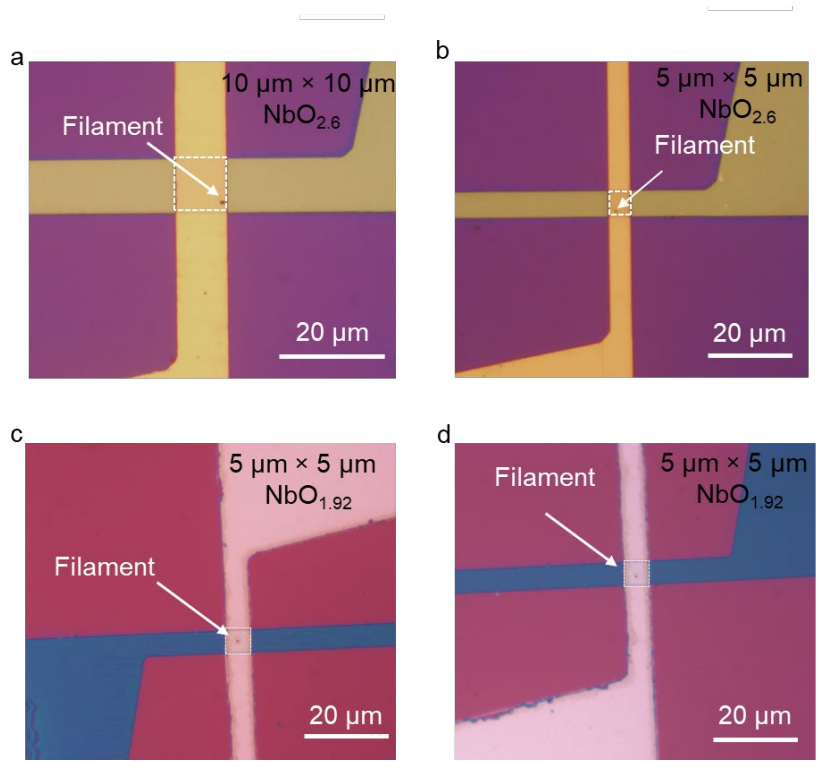

**Figure S2:** Optical images of the post formed devices.

# Rutherford backscattering spectroscopy of NbO$_x$ films

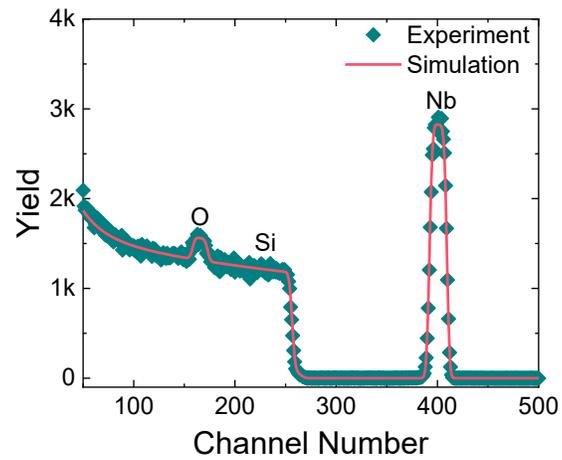

**Figure S3:** Representative Rutherford backscattered spectrum obtained from a DC reactively sputtered NbO$_x$ thin film deposited on Si substrate at Ar/O (19 sccm/1 sccm) atmosphere.

**Electroforming and subsequent NDR characteristics**

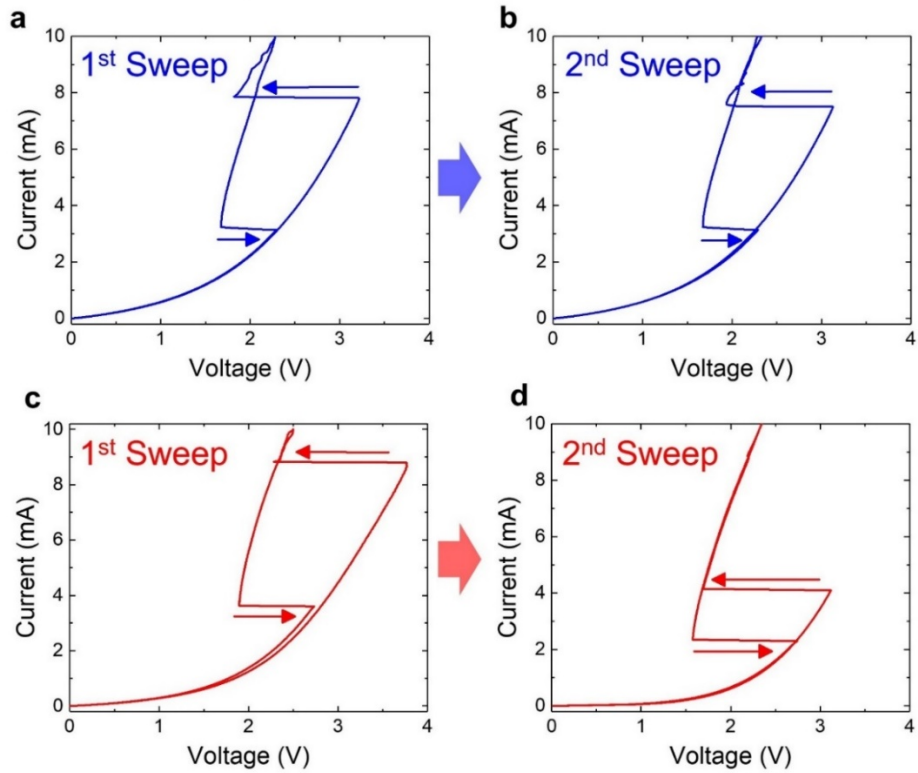

**Figure S4: Electroforming characteristics of a 5 µm × 5 µm cross-point device with high conducting film. a)** Electroforming and **b)** subsequent switching shows transient filament. **c)** Electroforming and d) subsequent characteristics indicates permanent filament formation during electroforming process.

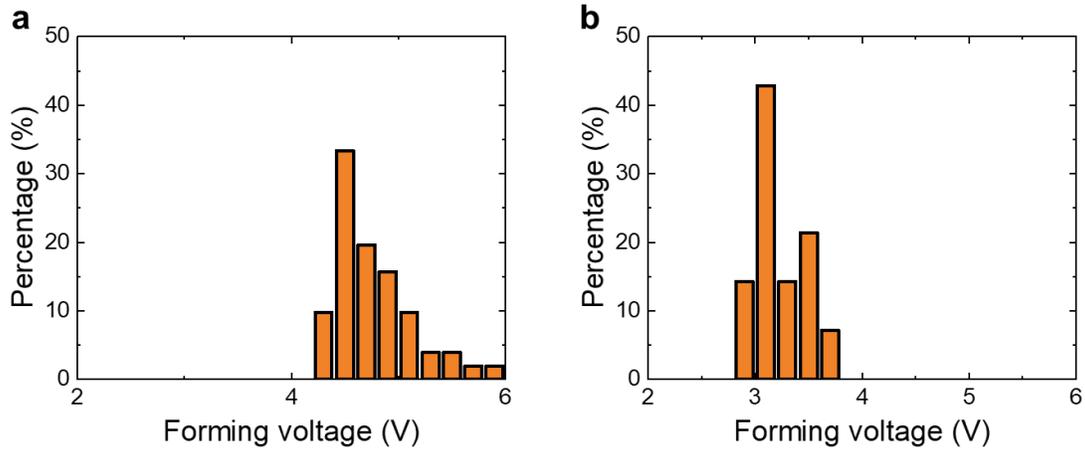

**Figure S5: Forming voltage distribution.** Histogram of forming voltage of **a)** stoichiometric NbOx (x=2.6) and **b)** substoichiometric NbOx (x=1.92).

**Finite Element Modelling**

Finite element modelling was performed with the COMSOL package. The model was based on a two-dimensional axisymmetric model of a Pt/NbO$_x$/Pt/SiO$_2$ device, as shown in Figure S6, and used the heat transfer and electric current modules to self consistently solve the heat and current continuity equations. The boundary conditions allowed radiative energy loss from the top surface of the device with an emissivity of 0.02, and set the outer edge of the device and the bottom of the SiO$_2$ layer at a fixed ambient temperature of 293 K, as indicated in Figure Sx. Simulations were performed as a function of device current with the current source applied to the top Pt electrode and the bottom Pt electrode serving as the electrical ground.

Material properties of Pt and SiO$_2$ were taken from COMSOL material library. The electrical conductivity of the NbO$_x$ layer was assumed to be governed by Poole-Frenkel conduction, as defined by Eq. S1. Relevant NbO$_x$ parameters are given in Table S2.

$$\sigma = \sigma_0 . exp[(-E_a + \sqrt{\beta\varepsilon})/kT] \qquad \text{Eq. S1}$$

where $\beta = \frac{e^3}{\pi \epsilon_0 \epsilon_r}$

Table S2: Properties of NbO$_x$ layer used in the simulation.

| NbO$_x$ Properties | Values |
|---|---|
| Thermal conductivity | 1.0 W/(mK) [31] |
| Activation energy | 0.45 eV [32-33] |
| $\sigma_o$ | 1x10$^5$ S/m |
| $\epsilon_r$ | 45 [32] |

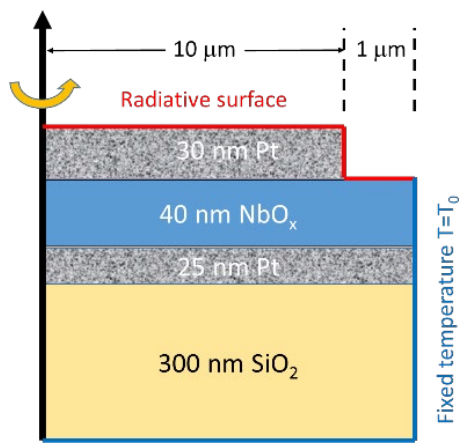

Figure S6: Schematic of the axisymmetric device structure employed for finite element simulations, including boundary conditions.

# Temperature mapping with thermoreflectance

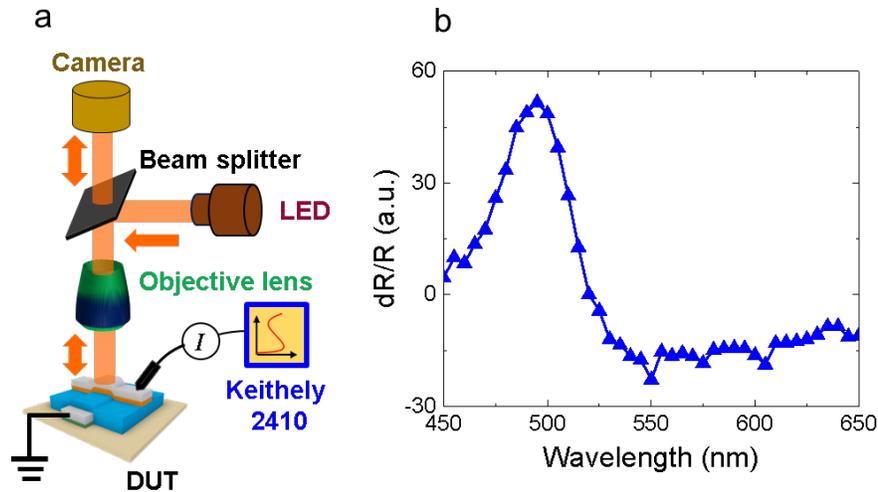

**Figure S7: Thermoreflectance measurement setup. a**, A simple representation of the thermoreflectance measurement system with an objective lenses with illumination 490 nm. **b**, Thermoreflectance co-efficient of gold (Au) surface under different illumination wavelengths.

# Temperature mapping with infrared microscopy

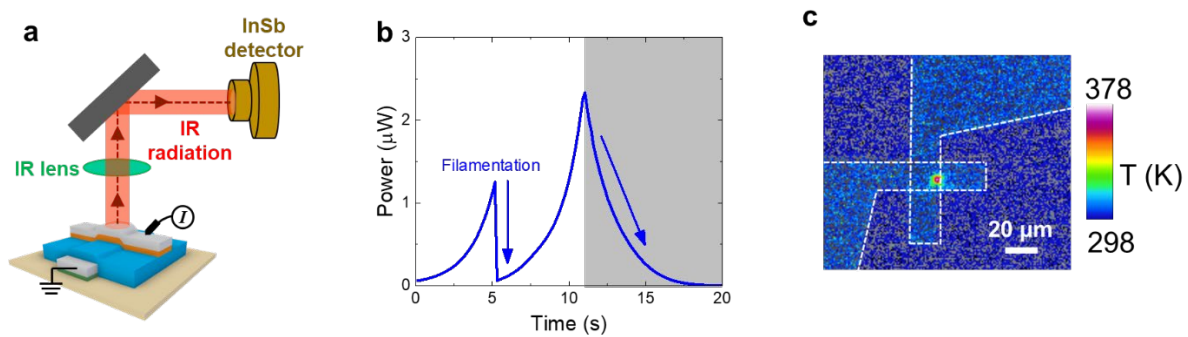

**Figure S8: Infrared microscopy measurement setup**. **a)** A simple schematic of the setup for thermal analysis for detecting the hot spot in 25 nm Au/20 nm Nb/45 nm NbO$_{2.60}$/25 nm Pt. **b)** In-situ electroforming power as a function of time. The shaded region is indicating reverse sweep during electroforming process. **c)** 2D plot of the temperature distribution at 3 mA of the top surface of the electrode. Further experimental details can be found elsewhere [34].